\documentclass[final,journal]{IEEEtran}

\usepackage{adjustbox}

\usepackage{array}
\usepackage{graphicx}
\usepackage{hyperref}
\usepackage{booktabs}
\usepackage{amsmath}
\usepackage{amssymb}
\newcommand{\norm}[1]{\left\lVert#1\right\rVert}
\usepackage{float}
\usepackage[dvipsnames]{xcolor}
\usepackage[square,numbers]{natbib}
\usepackage{caption}
\usepackage{subcaption}

\begin{document}

\title{Optimal Lane-Free Crossing of CAVs through Intersections}

\author{
    \IEEEauthorblockN{Mahdi Amouzadi, Mobolaji Olawumi Orisatoki and Arash M. Dizqah,~\IEEEmembership{Member,~IEEE}}\\
    \IEEEauthorblockA{Smart Vehicles Control Laboratory (SVeCLab), University of Sussex, Brighton BN1 9RH, UK
    \\\{m.amouzadi, a.m.dizqah\}@sussex.ac.uk}
    \thanks{Manuscript received , 2021; revised . 
    Corresponding author: Mahdi Amouzadi (email: m.amouzadi@sussex.ac.uk).}
}

\IEEEtitleabstractindextext{%
\begin{abstract}
Connected and autonomous vehicles (CAVs), unlike conventional cars, will utilise the whole space of intersections and cross in a lane-free order. This paper formulates such a lane-free crossing of intersections as a multi-objective optimal control problem (OCP) that minimises the overall crossing time, as well as the energy consumption of CAVs. The proposed OCP is convexified by applying the dual problem theory to the constraints that avoid collision of vehicles with each other and with road boundaries. The resulting OCP is smooth and solvable by gradient-based algorithms. Simulation results show that the proposed algorithm reduces the crossing time by an average of 40\% and 41\% as compared to, respectively, the state-of-the-art reservation-based and lane-free methods, whilst consuming the same amount of energy. Furthermore, it is shown that the resulting crossing time of the proposed algorithm is i) fixed to a constant value regardless of the number of CAVs, and ii) very close to its theoretical limit.\end{abstract}

\begin{IEEEkeywords}
signal-free intersection; path planning; connected and autonomous vehicles; dual problem theory
\end{IEEEkeywords}}

\maketitle

\IEEEdisplaynontitleabstractindextext

%
\IEEEpeerreviewmaketitle

\section{Introduction}
%
%
%
%
\IEEEPARstart{I}{ntersections} with traffic lights are inefficiently scheduled due to the limitations of human drivers which results in major congestion in urban traffic systems. Intersections also account for a large portion of all accidents (e.g. $47\%$ in the United State in 2010 \cite{national}). CAVs will be able to execute more complex manoeuvres than human drivers to cross intersections faster and safer, that also reduces energy consumption and increases traffic throughput \cite{rios2016survey}. However, these potential improvements require addressing three main challenges: collision avoidance, finding the minimum-time optimal solution instead of only checking the feasibility, and real-time implementation. Table \ref{lit_summary} provides a summary of challenges and the corresponding techniques.

Previous studies proposed three approaches to ensure collision avoidance among CAVs: 

i) Reserving the whole intersection for one of the CAVs at a time; The authors in \cite{zhang2017decentralized} designed an algorithm based on solving an OCP that jointly improves energy consumption and passenger comfort. The proposed OCP includes collision avoidance constraints that enforce CAVs to reserve the whole intersection for a period of time. A similar work is presented by Tallapragada et al. in \cite{tallapragada2019hierarchical}, where CAVs are split into clusters and each cluster reserves the whole area of the intersection for some time. The authors in \cite{xu2018distributed,di2019design} introduced a scheduling method where CAVs are placed into a virtual lane based on their distance to the centre of the intersection and their risk of collision. Then, crossing time of the intersection is scheduled between the CAVs in the virtual lane. The study in \cite{makarem2012fluent} also proposes an algorithm for CAVs based on the concept of reserving the whole intersection for one CAV at a time. It is shown that the proposed algorithm reduces the mean fuel consumption of every vehicle by 13.29–73.11\% as compared to traffic lights \cite{makarem2012fluent}.

ii) Reserving a finite number of specific points (called conflict points) instead of the whole of the intersection; Mirheli et al. \cite{mirheli2018development,mirheli2019consensus} designed 16 conflict points for a four-leg intersection. Each leg of the intersection includes exclusively left turn and straight lanes. The proposed algorithm enforces CAVs to reserve the approaching conflict point(s) prior to their arrival. A more recent study of the conflict-point-reservation technique is presented in \cite{xu2021comparison} where CAVs are capable of performing turning maneuvers. The algorithm initially finds the passing sequence of CAVs and then calculates the optimal control inputs analytically. The authors compared the energy consumption of CAVs when the algorithm finds the passing sequence using different strategies. It is shown that the best performance in terms of fuel consumption and travelling time is achieved when the passing sequence of CAVs is solved using the Monte Carlo Tree Search \cite{xu2021comparison}.

\renewcommand{\arraystretch}{1.5}
\begin{table*}[t]
\centering
\caption{Summary of the challenges of the intersection crossing problem and the corresponding techniques for the challenges} 
\label{lit_summary}
\begin{tabular}{l c c}
        \hline %
        \multicolumn{1}{m{6cm}}{\textbf{Challenge}} &\textbf{Techniques to address} & \textbf{Implemented in}\\
        \hline 
                            &Reservation of the whole Intersection & \cite{tallapragada2019hierarchical,xu2018distributed,pan2020optimal,zhang2017decentralized,di2019design,makarem2012fluent}\\\cline{2-3}
       Collision avoidance & {Reservation of conflict points}  & \cite{mirheli2018development,malikopoulos2021optimal,mirheli2019consensus}\\\cline{2-3}
                            & Lane-free  & \cite{li2018near,li2020autonomous}
        \\\hline
                            & Minimisation of fluctuation of the vehicles' acceleration & \cite{li2020autonomous,tallapragada2019hierarchical}\\\cline{2-3}
        \multicolumn{1}{m{6cm}}{Finding the minimum-time optimal solution instead of only checking the feasibility} & Minimisation of deviation from the speed limit  & \cite{liu2017distributed,hult2019optimal,hult2020optimisation,katriniok2017distributed}\\\cline{2-3}
                            & Minimisation of the crossing time & \cite{li2018near}
        \\\hline
                            & Centralised strategies with fully-observable data & \cite{li2018near,li2020autonomous,hult2019optimal,hult2020optimisation}\\\cline{2-3}
        Real-Time implementation & Decentralised strategies with fully connected CAVs & \cite{kloock2019distributed,malikopoulos2018decentralized}\\\cline{2-3}
                            & Decentralised strategies with partially connected CAVs  & \cite{bian2019cooperation,di2019design}
        \\\hline
\end{tabular}
\vspace{-10pt}
\end{table*}
\renewcommand{\arraystretch}{1}

iii) Utilising the whole space of intersections freely, a.k.a. lane-free crossing; Generally speaking, reservation-based collision avoidance approaches require CAVs following predefined paths and not fully exploiting the intersection area. These types of collision avoidance approaches are not efficient in terms of reducing travelling time and energy consumption. Prior studies developed lane-free algorithms based on OCP in \cite{li2020autonomous,li2018near}. To avoid collisions, the Euclidean distance between any pair of CAVs are constrained to be greater than a safe margin. This formulation of the collision avoidance constraints is non-convex \cite{colombo2012efficient}, and hence any optimisation problem including them are difficult to solve. Li et al. in \cite{li2018near} divided the non-convex problem of intersection crossing into two stages to make it tractable. At stage one, CAVs inform the central controller with their intention and then make a standard formation which is computed online. At stage two, the controller searches an offline constructed lookup table for the intended crossing scenario and finds the control inputs of each CAV. The authors suggested to solve offline an individual optimal control problem for any possible crossing scenario to construct the lookup table of the control commands. However, the resulting offline problems are still non-convex and solving them for all possible scenarios of 24 CAVs take around 358 years \cite{li2018near}. Alternatively, Li et al. in \cite{li2020autonomous} fixed the crossing time to a constant value and converted the minimum-time optimal control problem to a feasibility problem to solve online.

As the second challenge, the above-mentioned algorithms that only find a feasible (collision-free) solution to the problem of the intersection crossing do not fully exploit the CAVs' advantages to minimise the crossing time. In other words, minimising the crossing time is not part of their objectives. The studies carried out in \cite{tallapragada2019hierarchical} and \cite{li2020autonomous} focus on the passenger comfort and addressed the challenge by minimising fluctuation of the vehicles' acceleration. Other researchers in \cite{liu2017distributed,hult2019optimal,hult2020optimisation,katriniok2017distributed} optimised the motion of CAVs to move on the predefined paths with as close speed as possible to the speed limit of the intersection. These approaches formulate the intersection crossing as a simpler problem to solve rather than a minimum-time OCP. Moreover, the results show that these methods find the feasible solutions and only marginally improve the crossing time as compared to the traditional signalised intersections. The authors in \cite{li2018near} formulate the intersection crossing problem of CAVs as a minimum-time OCP to minimise the crossing time without any restrictions on the crossing paths (except the road boundaries). However, their algorithm is not time-effective for real-time applications. 

Finally, it is always challenging to implement optimal control strategies in real-time. CAVs are intelligent agents communicating to each other and to the infrastructure to share information such as location, speed, and intentions. The optimal strategies for crossing intersections, therefore, must operate on a network of cars (i.e., networked controller), considering the shared information to control multiple CAVs which are seeking conflicting objectives (like minimising their individual crossing times). A centralised topology with a fully available information of all the CAVs or different decentralised topologies, where the CAVs are fully or partially connected, can be used to calculate the optimal crossing strategy of all the CAVs. The centralised controllers receive information of all vehicles, compute trajectories, and send back the calculated trajectory of each individual CAV. There is no path planning at the CAV level and vehicles only follow the provided trajectories. Li et al. \cite{li2018near,li2020autonomous} proposed a centralised, but computationally expensive, optimal controller for multiple CAVs crossing a lane-free intersection. The centralised algorithms in \cite{hult2019optimal,hult2020optimisation} split the problem into two stages, finding the crossing order and calculating the control inputs to follow the attained crossing orders, to make it computationally tractable. 

In the decentralised strategies, on the other hand, each CAV computes its own trajectory by solving or approximating the solution of an optimal control problem to achieve a level of both the local and global objectives. The authors in \cite{malikopoulos2018decentralized,kloock2019distributed} formulated a decentralised OCP controller of the CAVs crossing intersections where each CAV has access to the shared information of all the others. Although decentralised algorithms find sub-optimal solutions, it is shown to be less computationally expensive as compared to the centralised counterparts \cite{malikopoulos2018decentralized,kloock2019distributed}.

However, the CAVs which are crossing an intersection cannot be practically fully connected to each other at all the times. This means that at any instance of time, each CAV only communicates with a subset of the others, i.e. partially connected. Bian et al. \cite{bian2019cooperation} proposed a framework where the CAVs travelling on the same lane can communicate to each other, but they estimate the states of the other not-connected vehicles. Reference \cite{di2019design} proposes a partially connected distributed algorithm based on the concept of virtual platooning. CAVs, first, form a virtual platoon and then optimise their arriving time to the intersection to avoid collision. This is a decentralised reservation-based algorithm that allows only one CAV  at a time within the intersection. Generally speaking, unlike the centralised controllers which are capable of finding the global optimum solution, decentralised controllers can only find sub-optimal strategies  \cite{elliott2019recent}.

In summary, majority of the literature propose reservation-based algorithms which calculate the feasible collision-free trajectory. To the best of the authors' knowledge, there is a limited number of literature dealing with lane-free minimum-time crossing of intersections. In addition, there is no analysis and comparison of lane-free intersections in terms of crossing time, energy consumption and passenger comfort. The lane-free minimum-time crossing is a non-convex problem and the current state-of-the-art approaches are computationally expensive. This paper addresses these gaps by the following novel contributions:

\begin{itemize}
\item Formulation of the lane-free crossing of CAVs through signal-free intersections as a minimum-time OCP;
\item Smoothing and convexification of the constraints that avoid collisions of CAVs with each other and with road boundaries. The constraints are replaced with the dual optimisation problem of their relaxed sufficient conditions;
\item Minimisation of the crossing time of multiple CAVs passing through intersections in a lane-free order. It is shown that the minimum crossing time calculated by the proposed algorithm is very close to its theoretical limit. The calculated optimal crossing time for a junction is fixed to a constant value regardless of the number of CAVs until reaching the maximum temporal-spatial capacity of the intersection;
\item Analysis and comparison of crossing time, energy consumption and passenger comfort of the proposed lane-free algorithm against a reservation-based method and a lane-free method. It is shown that the proposed lane-free algorithm significantly improves the crossing time and passenger comfort while consuming the same amount of energy as both the benchmark methods.
\end{itemize}

\begin{figure}[t]
    \centering
	\includegraphics[scale=0.35]{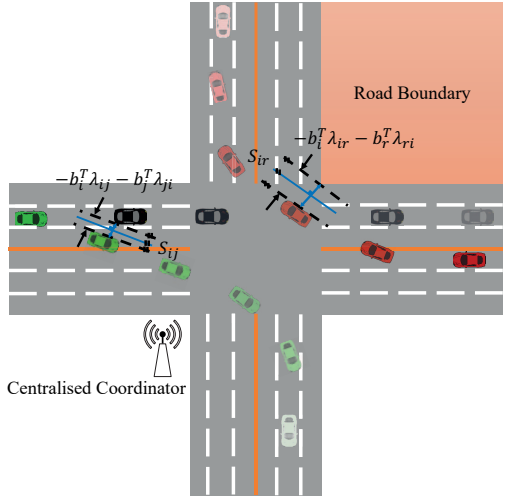}
	\caption{A lane-free and signal-free intersection. $S_{ij}$ and $S_{ir}$ are the separating hyperplanes between, respectively, two CAVs and a CAV and road boundary. $-b_{i}^\top\lambda_{ij} - b_{j}^\top\lambda_{ji}$ and $-b_{i}^\top\lambda_{ir} - b_{r}^\top\lambda_{ri}$ are distances which are formulated from the equivalent dual problem of the obstacle avoidance constraints. The dashed lines are the supporting hyperplanes.}
	\label{Intersection_layout}
	\vspace{-10pt}
\end{figure}

The remainder of this paper is organised as follows: Section \ref{system} describes the system of multiple CAVs crossing an intersection and presents the notations used throughout the paper; Section \ref{PF} formulates the lane-free crossing of CAVs through an intersection as a minimum-time optimal control problem; Section \ref{Discussion} provides numerical results obtained from simulation along with discussions, and Section \ref{conclusion} concludes the outcomes.

\section{System Description}\label{system}
\subsection{Lane-Free and Signal-Free Intersections}
Fig. \ref{Intersection_layout} illustrates an example layout of a lane-free and signal-free intersection. The figure includes three CAVs which are moving from their initial points, depicted with the most solid colour, towards their intended destinations which are with the most transparent colour. The intersection comprises of four approaches, each of them has a separate incoming and outgoing lane. In a lane-free intersection, vehicles can freely change their lanes in favour of faster crossing through the intersection. For instance, Fig. \ref{Intersection_layout} shows the green CAV overtakes the black CAV by using the opposite lane. It is worth noting that the intersection does not have traffic lights because the CAVs can directly communicate their states and intentions. In addition, all vehicles are of CAV type and no human-driven vehicle nor pedestrian are considered. In this study, there is a coordinator that receives all the information from the CAVs and centrally control them to efficiently and safely cross the intersection. This fully-connected and centralised topology provides the optimal solution as the benchmark for testing any decentralised strategies.

The black and green CAVs also show the collision avoidance. In this regard, the expression $-b_{i}^\top\lambda_{ij} - b_{j}^\top\lambda_{ji}$ is the dual representation of the distance between a pair of CAVs and $S_{ij}$ is the separating hyperplane placed between them. Similarly, the red CAV and the highlighted road boundary show the road boundary avoidance. $-b_{i}^\top\lambda_{ir} - b_{r}^\top\lambda_{ri}$ is the dual representation of distance between a CAV and a road boundary and $S_{ir}$ is the separating hyperplane between them. These equations are further explained in section \ref{PF}.  

\subsection{Vehicle Kinematics}
This study represents the lateral behaviour of CAVs with the bicycle model \cite{milliken1995race}. The bicycle model consists of two degree-of-freedom (DoF) which are sideslip angle $\beta_{i}$ and yaw rate $r_{i}$, as in Fig. \ref{Vehicle_model}. The model also includes an additional DoF for the longitudinal velocity $V_{i}$. The equations of these DoFs along with other three to model the ground-fixed location, construct a set of differential equations to represent $\text{CAV}_{i}$ as follows:

\vspace{-10pt}
\begin{align}
    \frac{d}{dt} \begin{bmatrix}\notag
    r_{i}\\[0.1cm]
    \beta_{i}\\[0.1cm]
    V_{i}\\[0.1cm]
    x_{i}\\[0.1cm]
    y_{i}\\[0.1cm]
    \theta_{i}\\[0.1cm]
    \end{bmatrix}(t)
    =
    &\begin{bmatrix}
    \frac{\tilde{N_{r}}}{I_{z}\cdot V_{i}(t)} \cdot r_{i}(t) + \frac{N_{\beta}}{I_{z}} \cdot \beta_{i}(t)\\[0.1cm]
    (\frac{\tilde{Y_{r}}}{m \cdot V_{i}(t)^2} - 1) \cdot r_{i}(t) + \frac{Y_{\beta}}{m\cdot V_{i}(t)} \cdot \beta_{i}(t) \\[0.1cm]
     0\\[0.1cm]
    V_{i}(t) \cdot cos\theta_{i}(t)\\[0.1cm]
    V_{i}(t) \cdot sin\theta_{i}(t)\\[0.1cm]
    r_{i}(t)\\[0.1cm]
    \end{bmatrix}
    +\\ 
    &\begin{bmatrix}
    0 & \frac{N_{\delta}}{I_{z}} \\ 
    0 & \frac{Y_{\delta}}{m \cdot V_{i}(t)}  \\ 
    1 & 0 \\
    0 & 0 \\
    0 & 0 \\ 
    0 & 0 \\ 
    \end{bmatrix}
    \begin{bmatrix}
    a_{i}\\
    \delta_{i}
    \end{bmatrix}(t)
    ,
    t \in [t_{0},t_{f}].
    \label{vehilce_model}
\end{align}
where $\textbf{x} = [r_{i},\beta_{i},V_{i},x_{i},y_{i},\theta_{i}]^T$ and $\textbf{u} =[a_{i}, \delta_{i}]^T$ are, respectively, the system states and control inputs of $\text{CAV}_i$. $\textbf{z}_{i}=[x_{i},y_{i},\theta_{i}]^T$ refers to the pose of $\text{CAV}_i$ in non-inertial reference system. $a_{i}(t)$ and $\delta_{i}(t)$ are, respectively, the acceleration ($m^2$) and steering angle ($rad$) of the vehicle. The constants $m$ and $I_{z}$ denote mass ($kg$) and moment of inertia ($kg.m^2$) of the vehicle. $t_{0}$ and $t_{f}$ represent the starting and final time ($s$) of crossing the intersection. The vehicle parameters $\tilde{N_{r}}$, $N_\beta$, $N_\delta$, $\tilde{Y_{r}}$, $Y_\beta$ and $Y_\delta$ are calculated as follows \cite{milliken1995race}:

\vspace{-10pt}
\begin{align*}\label{diff}
    \tilde{N_{r}} &= l_{f}^2 \cdot C_{F} + l_{r}^2 \cdot C_{R},\notag\\
    N_{\beta} &= l_{f} \cdot C_{F} - l_{r} \cdot C_{R},\notag\\
    N_{\delta} &= -l_{f} \cdot C_{F},\notag\\
    \tilde{Y_{r}} &=l_{f} \cdot C_{F} - l_{r} \cdot C_{R},\notag\\
    Y_{\beta} &= C_{F} + C_{R},\notag\\
    Y_{\delta} &= - C_{F}.\notag\\
\end{align*}
where $C_{F}$ and $C_{R}$ are, respectively, the cornering stiffness of the front and rear tyres. $l_{f}$ and $l_{r}$ are distance of the front and rear axis from center of gravity of the vehicle.

To ensure CAVs drive within their dynamic limitations, the following constraints are enforced for each $\text{CAV}_i$:

\vspace{-10pt}
\begin{subequations}\label{Limits}
    \begin{align}
        \underline{V} \leq V_{i}(t)\leq \bar{V},\\
        \underline{a} \leq \rvert a_{i}(t) \rvert \leq \bar{a},\\
        \underline{\delta} \leq \rvert \delta_{i}(t) \rvert \leq \bar{\delta},\\
        \underline{r} \leq \rvert r_{i}(t) \rvert \leq \bar{r},\\
        \underline{\beta} \leq \rvert \beta_{i}(t) \rvert \leq \bar{\beta}.
     \end{align}
\end{subequations}
where $\overline{.}$ and $\underbar{.}$ are, respectively, the upper and lower boundaries.

\subsection{Polytopic Representation of CAVs and Road Boundaries}

This study represents each $\text{CAV}_i$, when $i \in \{1..N\}$ and $N$ is the total number of CAVs, as a rectangular polytope $\tilde{\beta}_{i}$ (i.e., a convex set) that is the intersection area of half-space linear inequalities $\tilde{A}_{i}X \le \tilde{b}_{i}$ at the origin, where $X=[x,y]^T$ is a Cartesian point. In this paper, all CAVs have the same size which are defined with:
\begin{equation}
\tilde{A}_{i} = \left[ \begin{smallmatrix}
  1 & -1 & 0 & 0  \\
  0 & 0 & -1 & 1 
\end{smallmatrix} \right]^\top,\quad 
\tilde{b} =[l/2, l/2, d/2, d/2]^\top.
\end{equation}

\begin{flushleft}where $l$ and $d$ denote, respectively, the wheelbase and track of CAVs.\end{flushleft} 

As $\text{CAV}_i$ moves to a new pose $\textbf{z}_i(t) = [x_i(t), y_i(t), \theta_i(t)]^T$, the original polytope $\tilde{\beta}_{i}$ of the CAV is transformed to $\beta_{i}$ as follows:

\vspace{-10pt}
\begin{align}
\label{transfor}
(\tilde{\beta_{i}}, \textbf{z}_i(t)) \mapsto \beta_{i}(\textbf{z}_i;t): A_{i}(\textbf{z}_i;t)X(t) \le b_{i}(\textbf{z}_i;t).
\end{align}

\begin{figure}[t]
    \centering
    \begin{subfigure}[t]{0.35\textwidth}
         \centering
         \includegraphics[scale=0.45]{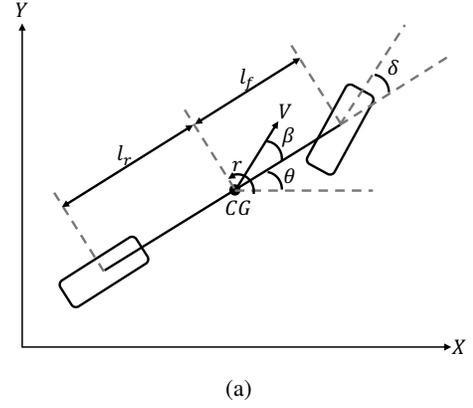}
         \caption{}
         \label{Vehicle_model}
         \vspace{15pt}
     \end{subfigure}
     \begin{subfigure}[t]{0.4\textwidth}
         \centering
         \includegraphics[scale=0.55]{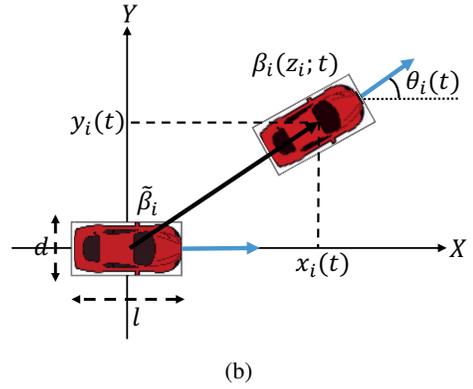}
         \caption{}
         \label{Transformation}
    \end{subfigure}
    \caption{(a) The bicycle model of vehicles. (b) Transformation of each $\text{CAV}_i$ from $\tilde{\beta}_{i}$ to $\beta_{i}(\textbf{z}_i;t)$ where $\textbf{z}_i(t) = [x_i(t), y_i(t),\theta_i(t)]^T$.}%
    \label{ModelTransformation}%
    \vspace{-10pt}
\end{figure}

\begin{flushleft}where:\end{flushleft}
\vspace{-7pt}
\begin{subequations}\label{transformation}
\begin{align}
&{A}_{i}(z_i;t) = \tilde{A}_{i} \left[ \begin{smallmatrix}
  cos\theta_{i}(t) & sin\theta_{i}(t) \\
  -sin\theta_{i}(t) & cos\theta_{i}(t) 
\end{smallmatrix} \right],
 \\
b_{i}(z_i;t) &= \tilde{\beta}_{i} + \tilde{A}_{i} \left[ \begin{smallmatrix}
  cos\theta_{i}(t) & sin\theta_{i}(t) \\
  -sin\theta_{i}(t) & cos\theta_{i}(t) 
\end{smallmatrix} \right] [x_{i}(t), y_{i}(t)]^\top.
\end{align}
\end{subequations}
Fig. \ref{Transformation} provides a graphical representation of (\ref{transfor}).

Road boundaries are also modelled as convex polytopic sets $O_r$, when $r \in \{1..N_{r}\}$ and $N_{r}$ is the total number of road boundaries which is 4 for four-legged intersections. 

Based on these representations, there is no collision between $\text{CAV}_i$ and $\text{CAV}_j$ iff $\beta_{i}(\textbf{z}_i;t) \cap \beta_{j}(\textbf{z}_j;t) = \emptyset, \forall t \in [t_0,t_f]$. Similarly, CAVs do not collide with road boundaries when the intersection of their sets is always empty, i.e. $\beta_{i}(\textbf{z}_i;t) \cap O_{r} = \emptyset, \forall t \in [t_0,t_f]$.

\section{Problem Formulation}\label{PF}
This section formulates simultaneous crossing of multiple CAVs through a lane-free and signal-free intersection as an optimal control problem. The formulated OCP minimises the overall crossing time as well as the energy consumption of CAVs whilst avoiding collisions of each vehicle with others and with road boundaries. Rest of the sections provide collision avoidance constraints, initial and terminal conditions and the objective function before presenting the overall OCP formulation.

\subsection{Constraints to Avoid Collisions Between CAVs}\label{CAVs_CAC}
To avoid collisions between any $\text{CAV}_i$ and $\text{CAV}_j$ $\forall i\neq j \in \{1..N\}$, their polytopic sets should not intersect, i.e. $\beta_{i} \cap \beta_{j} = \emptyset$ where $\beta_{i} = \{\text{X} \in \mathbb{R}^{2}|A_{i}\text{X} \leq b_{i}\}$ and $\beta_{j} = \{\text{Y} \in \mathbb{R}^{2}|A_{j}\text{Y} \leq b_{j}\}$. However, these are non-convex and non-differentiable conditions and enforcing them as constraints in an OCP will make the problem non-convex and difficult solve. To preserve differentiability and continuity, $\beta_{i} \cap \beta_{j} = \emptyset$ is replaced by the following sufficient condition which has negligible effect on the optimality of the solution for small values of $d_{min}$ \cite{zhang2020optimization}: 

\vspace{-10pt}
\begin{align}\label{primal}
    dist (\beta_{i},\beta_{j}) =\underset{\text{X,Y}}{\text{min}}\{\norm{\text{X-Y}}_{2}\;|\;A_{i}\text{X}&\leq b_{i},\; A_{j}\text{Y}\leq b_{j}\} \geq d_{min};\notag \\ 
    &\forall i\neq j \in \{1..N\}.
\end{align}
where $d_{min}$ is the minimum safe distance between CAVs.

It is known that the problem of finding the minimum distance between a polytope $\beta_{i}$ and another given polytope $\beta_{j}$ is convex \cite{boyd_vandenberghe_2004}. Therefore the substituted sufficient condition, $dist (\beta_{i},\beta_{j}) \geq d_{min}$, is convex and since $\beta_{j}$ is not an empty set, the strong duality holds \cite{zhang2020optimization}. This means that the solution of the primal problem $dist (\beta_{i},\beta_{j}) \geq d_{min}$ is the same as the one of its dual problem which is as follows:

\vspace{-10pt}
\begin{align}\label{dual}
  \text {dist} (\beta_{i},\beta_{j}) :=\:&\underset{\lambda_{ij},\lambda_{ji},s_{ij}}{\text{max}} -b_{i}^\top\lambda_{ij} - b_{j}^\top\lambda_{ji} \\
&\text{s.t.} \notag \quad A_{i}^\top\lambda_{ij} + s_{ij} = 0, A_{j}^\top\lambda_{ji} - s_{ij} = 0, \notag\\ 
&\quad \quad  \norm{s_{ij}}_{2} \leq 1, -\lambda_{ij} \leq 0, -\lambda_{ji} \leq 0; \notag\\
& \quad \quad  \forall i\neq j \in \{1..N\}.\notag
\end{align}
where $\lambda_{ij}$, $\lambda_{ji}$, and $s_{ij}$ are the dual variables and $A_i$ and $b_i$ are as in (\ref{transformation}) (the deviation of dual problem (\ref{dual}) from primal problem (\ref{primal}) is shown in \cite{firoozi2020distributed}). 

Combining (\ref{dual}) with (\ref{primal}), the objective function of (\ref{dual}) subject to its constraints must be greater than or equal to $d_{min}$ in order to avoid collisions. However, (\ref{dual}) can be substituted by $\{\exists \lambda_{ij} \geq 0, \lambda_{ji} \geq 0, s_{ij} : -b_{i}^\top\lambda_{ij} - b_{j}^\top\lambda_{ji} \geq d_{cmin}, A_{i}^\top\lambda_{ij} + s_{ij} = 0, A_{j}^\top\lambda_{ji} - s_{ij} = 0, \norm{s_{ij}}_{2} \leq 1\}$ because the existence of a feasible solution $\lambda_{ij,feas}$, $\lambda_{ji,feas}$, and $s_{ij,feas}$ where $-b_{i}^\top\lambda_{ij,feas} - b_{j}^\top\lambda_{ji,feas} \geq d_{min}$ is a sufficient condition to ensure $ dist (\beta_{i},\beta_{j}) \geq d_{min}$, i.e. to avoid collisions \cite{firoozi2020distributed}. As seen in Fig. \ref{Intersection_layout}, $S_{ij}$ is a separating hyperplane between $\text{CAV}_i$ and $\text{CAV}_j$ and $S_{ij}$ = $S_{ji}$.

\subsection{Constraints to Avoid Collisions with Road Boundaries}\label{CAVs_ROAD}
Each $\text{CAV}_{i}$ must also avoid all the road boundaries, i.e. $\beta_{i} \cap O_{r} = \emptyset$ where $\beta_{i} = \{\text{X} \in \mathbb{R}^{2}|A_{i}\text{X} \leq b_{i}\}$ and $O_{r} =\{\text{Y} \in \mathbb{R}^{2}|A_{r}\text{Y} \leq b_{r}\}$. Similar to section \ref{CAVs_CAC} collision avoidance between CAVs, $\beta_{i} \cap O_{r} = \emptyset$ is replaced by the following sufficient condition:

\vspace{-10pt}
\begin{align}\label{primal_rd}
    \text {dist} (\beta_{i},O_{r}) =\underset{\text{X,Y}}{\text{min}}\{\norm{\text{X-Y}}_{2}|A_{i}\text{X}\leq &b_{i},A_{r}\text{Y}\leq b_{r}\} \geq d_{rmin}; \notag \\
    &\forall r \in \{1..N_{r}\}.
\end{align}
where $d_{rmin}$ is the minimum safety distance between CAVs and road boundaries.

The dual problem of (\ref{primal_rd}) is then substituted with the sufficient condition $\{\exists \lambda_{ir} \geq 0, \lambda_{ri} \geq 0, s_{ir} : -b_{i}^\top\lambda_{ir} - b_{r}^\top\lambda_{ri} \geq d_{rmin}, A_{i}^\top\lambda_{ir} + s_{ir} = 0, A_{j}^\top\lambda_{ri} - s_{ir} = 0, \norm{s_{ir}}_{2} \leq 1\}$
where $\lambda_{ir},\lambda_{ri}$, and $s_{ir}$ are the dual variables. $S_{ir}$ is the separating hyperplane between CAVs and road boundaries (see Fig. \ref{Intersection_layout}). 

\subsection{Objective Function}
\label{objectiveFunction}
CAVs are expected to reach their terminal pose as fast as possible while consume energy as little as possible. Therefore, this paper proposes the objective function (\ref{cost}) that minimises the overall crossing time of all CAVs and the error between the current and final pose, as well as the energy consumption of each vehicle:
\begin{align}\label{cost}
   & J(\textbf{z}_1(.),..,\textbf{z}_N(.)) = \alpha(t_{f} - t_{0})^2 +\\
    &\int_{t_{0}}^{t_{f}} \sum_{i=1}^{N}\,[(\textbf{z}_{i}(t)-\textbf{z}_{i}(t_{f}))^\top \textbf{Q} (\textbf{z}_{i}(t)- \textbf{z}_i(t_{f})) + \gamma a_{i}(t)^2]\: dt \: . \nonumber
\end{align}
\\
where $\alpha$, $\textbf{Q}$ and $\gamma$ are the gain factors related to the crossing time, CAVs' pose and energy consumption respectively. The gains are selected based on trail and error to best normalise the cost function. The expression $(t_{f}-t_{0})^2$ minimises the crossing time of all CAVs. The Lagrange term penalises the error between the current pose $\textbf{z}_{i}(t)$ and the final pose $\textbf{z}_{i}(t_{f})$ as well as the acceleration $a_{i}(t)^2$ of vehicles. The final pose of CAVs $\textbf{z}_{i}(t_{f})$ is directly imposed in the objective function and indicates the intended destination of each $\text{CAV}_{i}$.

\subsection{Optimal Control Problem}
Lane-free crossing of multiple CAVs through a signal-free intersection is formulated as the following optimal control problem:

\vspace{-10pt}
\begin{subequations} \label{OCPV1}
\begin{align}
&\{a_i(.),\delta_i(.)\}^* = \\
&\quad \quad \text{arg}\:\underset{\substack{t_{f},a_i(.),\delta_i(.)}}{\text{min }} J(\textbf{z}_1(.),..,\textbf{z}_N(.)) := (\ref{cost}), \\
&\qquad\text{s.t.}  \quad (\ref{vehilce_model}),\; (\ref{Limits}) \label{kin},\\
&\qquad \qquad\beta_{i}(t) \cap \beta_{j}(t) = \emptyset \label{colli};\: \forall i\neq j \in \{1..N\},\\
&\qquad \qquad\beta_{i}(t) \cap O_{r}(t) = \emptyset \label{colli_r};\: \forall i \in \{1..N\},\notag\\
&\qquad\qquad\qquad\qquad\qquad\qquad\forall r \in \{1..N_{r}\},\\
&\qquad \qquad \textbf{z}_i(t_{0}) = \textbf{z}_{i,0}, \: \forall i \in \{1..N\}, \: t \in[t_0, t_f].\notag
\end{align}
\end{subequations}
where (\ref{kin}) refers to the vehicle kinematics and (\ref{colli}) and (\ref{colli_r}) denote, respectively, collision avoidance constraints of each CAV with others and with road boundaries.

As discussed in sections \ref{CAVs_CAC} and \ref{CAVs_ROAD}, the non-differentiable and non-convex collision avoidance constraints (\ref{colli}) and (\ref{colli_r}) are substituted by the dual problem of their sufficient conditions (\ref{primal}) and (\ref{primal_rd}), and then (\ref{OCPV1}) is reformulated as the following smooth and continuous problem, which is solvable by the state-of-the-art gradient-based algorithms:

\vspace{-10pt}
\begin{subequations} \label{OCP}
\begin{alignat}{10}
&\{a_i(.),\delta_i(.)\}^* = \\ &\quad \quad\text{arg}\:\underset{\substack{t_{f},a_i(.),\delta_i(.)\\\lambda_{ij},\lambda_{ji},s_{ij},\\\lambda_{ri},\lambda_{ir},s_{ir}}}{\text{min}}  J(\textbf{z}_1(.),..,\textbf{z}_N(.)) := (\ref{cost}), \notag \\
 & \quad \quad \text{s.t.} \quad (\ref{vehilce_model}), (\ref{Limits}),\\
 &\qquad \qquad    -b_{i}(\textbf{z}_{i}(t))^\top\lambda_{ij}(t)-b_{j}(\textbf{z}_{j}(t))^\top \lambda_{ji}(t)\ge d_{min} \label{CA_A} \\
 &\qquad \qquad   A_{i}(\textbf{z}_{i}(t))^\top \lambda_{ij}(t)+s_{ij}(t)=0 \label{CA_B}\\
 &\qquad \qquad    A_{j}(\textbf{z}_{j}(t))^\top \lambda_{ji}(t)-s_{ij}(t)=0 \label{CA_C} \\
  & \qquad \qquad   -b_{i}(\textbf{z}_{i}(t))^\top\lambda_{ir}(t)-b_{r}^\top \lambda_{ri}(t)\ge d_{rmin} \label{RA_A}\\
 &\qquad \qquad    A_{i}(\textbf{z}_{i}(t))^\top \lambda_{ir}(t)+s_{ir}(t)=0 \label{RA_B}\\
 &\qquad \qquad    A_{r}^\top \lambda_{ri}(t)+s_{ir}(t)=0 \label{RA_C}\\
 &\qquad \qquad    \lambda_{ij}(t),\;\lambda_{ji}(t),\;\lambda_{ir}(t),\;\lambda_{ri}(t)\geq 0, \\ 
 & \qquad \qquad   \norm{s_{ij}(t)}_{2}\leq 1,\;\norm{s_{ir}(t)}_{2}\leq1,\\
 &  \qquad \qquad  \textbf{z}_i(t_{0}) = \textbf{z}_{i,0},\\
 &  \qquad \qquad   \forall i\neq j \in \{1..N\}, \forall r \in \{1..N_r\}. \notag
\end{alignat}
\end{subequations}
where $A_{i}$ and $b_{i}$ are functions of each CAV's pose $\textbf{z}_{i}(t)$, and present $\text{CAV}_{i}$ polytope at each time step $t$. Problem (\ref{OCP}) is solved at time $t_{0}$ for $N$ CAVs until the terminal time $t_{f}$. The solution to this problem is optimal trajectories of the control signals $a_i(.)^*$ and $\delta_i(.)^*$ of each $\text{CAV}_i$ for each $t\in[t_0,t_f]$, as well as a terminal time $t_{f}$. CAVs follow their calculated trajectories to arrive final destination at the terminal time $t_{f}$.
\\

\renewcommand{\arraystretch}{1.2}
\begin{table}[t]
    \caption{Main parameters of the model} 
    \label{settings} 
    \begin{tabular}{l l c}
    \toprule
    \textbf{Parameter(s)} & \textbf{Description}  & \textbf{Value(s)}\\ [0.5ex] 
    \midrule
    $\text{m}\: (kg)$ & \multicolumn{1}{m{4cm}}{mass of each CAV} & 1204  \\
    $\text{d}_{min}\: (m)$ & \multicolumn{1}{m{4cm}}{minimum distance between CAVs} & 0.1  \\
    $\text{d}_{rmin} (m)$ & \multicolumn{1}{m{4cm}}{minimum distance between CAVs and road boundaries} & 0.1 \\
    $d$ (-) & \multicolumn{1}{m{4cm}}{number of collocation points}  & 5  \\[0.5ex]
    $N_{p}\:$(-) & \multicolumn{1}{m{4cm}}{prediction horizon (prediction step is calculated)}  & 15  \\[0.5ex]
    $V_{max}\: (m/s)$ & \multicolumn{1}{m{4cm}}{bounds on $V_{i}$} & 25  \\[0.5ex]
    $\delta_{max}\:(rad)$ & \multicolumn{1}{m{4cm}}{bounds on $|\delta_{i}|$} & 0.67 \\[0.5ex]
    $a_{max}\:(m/s^2)$ & \multicolumn{1}{m{4cm}}{bounds on $|a_{i}|$} & 3 \\[0.5ex]
    $r_{i}$ (m) & \multicolumn{1}{m{4cm}}{bounds on $|r_{i}|$} & 0.7 \\[0.5ex]
    $\beta_{max}$ (rad/s) & \multicolumn{1}{m{4cm}}{bounds on $|\beta_{i}|$} & 0.5 \\[0.5ex]
    \multicolumn{1}{m{2.5cm}}{$V_i(t_0)\; \forall i \in \{1..N\}$ $\,(m/s)$}& initial speed of CAVs & 10 \\[0.5ex]
    \bottomrule
    \end{tabular}
\end{table}

The initial pose $\textbf{z}_i$, i.e. initial position, heading angle and initial speed $V_i$ of all $\text{CAV}_i \: \forall i \in \{1..N\}$ within the control zone are known. The remaining of the states and the initial inputs to the CAVs are also assumed as zero. These initial conditions at $t = t_{0}$ are feasible solutions of the OCP.  

\section{Results and Discussion}\label{Discussion}
In this section, performance of the proposed algorithm is compared against two state-of-the-art benchmarks in terms of crossing time, energy consumption and passenger comfort. For doing this, this study employs the intersection scenario proposed in \cite{malikopoulos2021optimal}, which is named test scenario one hereafter. The first benchmark is a conflict-point-reservation approach presented in \cite{malikopoulos2021optimal}, where each CAV calculates its own trajectory by jointly minimising the travelling time and energy consumption. The calculated reservation times  for each conflict point are then shared with other vehicles through a centralised coordinator. Vehicles entering the intersection later read these reserved times and treat them as additional collision avoidance constraints when they plan their own trajectory. The second benchmark is a lane-free method proposed in \cite{li2020autonomous} where CAVs can freely use all the space of junction, as long as there is no collision. The proposed algorithm in \cite{li2020autonomous} calculates the control inputs for a relatively large given value of crossing ti;me. 

The algorithms are compared within test scenario one in terms of crossing time, average and standard deviation of speed and energy consumption for different number of CAVs between 2 to 12. The initial and terminal pose of CAVs are chosen randomly and there exists at least one CAV performing a left-turn manoeuvre in each test. The proposed and benchmark algorithms share the same values for the starting and terminal positions of vehicles, the maximum permissible speed and acceleration of CAVs, which are provided in Table \ref{settings} along with other critical parameters.

Furthermore, the performance of the proposed algorithm is analysed for a more complex scenario, which is named test scenario two hereafter. This second test scenario involves up to 21 CAVs, and allows any travelling direction by CAVs (e.g., right, straight and left).

CasADi \cite{Andersson2019} with IPOPT \cite{wachter2006implementation} are used to solve the formulated nonlinear OCP in (\ref{OCP}). CasADi directly discretises continuous-time OCPs by a collocation method and constructs an equivalent nonlinear programming (NLP) \cite{rosmann2020time}. The resulting NLP is then solved using IPOPT which is an implementation of the interior-point method (IPM) \cite{wachter2006implementation}. To improve the computation time, IPOPT is linked to Intel® oneAPI Math Kernel Library (oneMKL, \color{blue} https://software.intel.com\color{black}), which includes high-performance implementation of the \textit{MA27} linear solver. All the results are calculated with MATLAB running on a Linux Ubuntu 20.04.0 LTS server with a 3.7 GHz Intel® Core i7 and 32~GB of memory.

\begin{table}[t]
    \caption{Performance of the proposed lane-free method for test scenario one as compared to the reservation-based method in \cite{malikopoulos2021optimal} and the lane-free method in \cite{li2020autonomous}.}
    \begin{tabular}{m{3.3cm} m{0.43cm} m{0.43cm} m{0.43cm} m{0.43cm} m{0.43cm} m{0.43cm} m{0.43cm}}
    \toprule
    \textbf{Number of CAVs $\rightarrow$} & \textbf{2} &\textbf{4} & \textbf{6} & \textbf{8} & \textbf{10} &\textbf{12} \\ 
    \hline
    \multicolumn{1}{m{3.3cm}}{\textbf{The proposed algorithm}} & & & & & & \\
    \multicolumn{1}{m{3.3cm}}{Crossing time (s)}  & 4.56 & 4.57 & 4.57   & 4.57   & 4.57   & 4.57\\ 
    \multicolumn{1}{m{3.3cm}}{Average speed (m/s)}  & 15.17 & 15.78 & 15.58   & 15.09   & 15.35   & 15.52\\
    \multicolumn{1}{m{3.3cm}}{Standard deviation of speed}  & 3.46 & 3.77 & 3.66   & 3.43   & 3.58   & 3.66\\
    \multicolumn{1}{m{3.3cm}}{Energy consumption (kWh)}  & 0.1&	0.23&	0.33&	0.39&	0.52&	0.65\\
    \multicolumn{1}{m{3.3cm}}{Travelled distance (m)}  & 130 & 270 & 400  &518  & 658   & 799\\
    \hline
    \multicolumn{1}{m{3.3cm}}{\textbf{Reservation-based \cite{malikopoulos2021optimal}}} & & & & & & \\
    \multicolumn{1}{m{3.3cm}}{Crossing time (s)}  & 6.29 & 6.29 & 12.50   & 12.93   & 12.92   & 12.92\\ 
    \multicolumn{1}{m{3.3cm}}{Average speed (m/s)}  & 12.69 & 13.53 & 11.29   & 9.91   & 10.59   & 11.10 \\
    \multicolumn{1}{m{3.3cm}}{Standard deviation of speed}  & 2.32 & 2.50 & 4.44   & 4.44   & 4.56   & 4.56\\
    \multicolumn{1}{m{3.3cm}}{Energy consumption (kWh)}  & 0.03	&0.07	&0.05	&0.04	&0.06	&0.07\\
    \multicolumn{1}{m{3.3cm}}{Travelled distance (m)}  & 134 & 265 & 406   & 507   & 630   & 750\\\hline
    \multicolumn{1}{m{3.3cm}}{\textbf{Lane-free \cite{li2020autonomous}}} & & & & & & \\
    \multicolumn{1}{m{3.3cm}}{Crossing time (s)}  & 10 & 10 & 10   & 10   & 10   & 10\\ 
    \multicolumn{1}{m{3.3cm}}{Average speed (m/s)}  & 10.11 & 10.04 & 9.98   & 10.02   & 10.07   & 10.03\\ 
    \multicolumn{1}{m{3.3cm}}{Standard deviation of speed}  & 0.14 & 0.12 & 0.14   & 0.17   & 0.23   & 0.21\\
    \multicolumn{1}{m{3.3cm}}{Energy consumption (kWh)}  & 0.001	&0.001	&0.001	&0.003	&0.006	&0.006\\
    \multicolumn{1}{m{3.3cm}}{Travelled distance (m)}  & 195 & 387 & 577   & 773   & 973   & 1160\\
    \bottomrule
    \end{tabular}\label{benchmark_results}
    \vspace{-2pt}
\end{table}

\begin{table}[t]
    \centering
    \caption{Simulation results of the proposed algorithm in test scenario two for different number of CAVs.}
        \begin{tabular}{m{2.75cm} m{0.38cm} m{0.38cm} m{0.38cm} m{0.38cm} m{0.38cm} m{0.38cm} m{0.38cm} m{0.38cm} m{0.38cm}}
        \toprule
        \textbf{Number of CAVs $\rightarrow$} & \textbf{3} & \textbf{6} & \textbf{9} & \textbf{12} & \textbf{15} & \textbf{18} & \textbf{21}  \\ \hline
        Crossing time (s)  & 4.57 & 4.57 & 4.57 & 4.56 & 4.57 & 4.58 & 4.57\\
        Average speed (m/s)  & 13.18 & 14.52 & 14.50 & 14.16 & 14.12 & 13.77 & 13.55\\
        Standard deviation of speed  & 3.94 & 4.01 & 3.81 & 3.87 & 4.13 & 4.18 & 4.07\\
        Energy Consumption (kWh)& 0.11&	0.27&	0.4	&0.5&	0.64&	0.7& 0.8\\
        Travelled distance (m)  & 170 & 373 & 561& 730 & 910& 1067 & 1224\\  
        \bottomrule
        \end{tabular}
        \label{complex_Ta}
    \vspace{-10pt}
\end{table}

\begin{figure*}[ht]
    \centering
    \begin{subfigure}[t]{0.49\textwidth}
         \centering
         \includegraphics[scale=0.33]{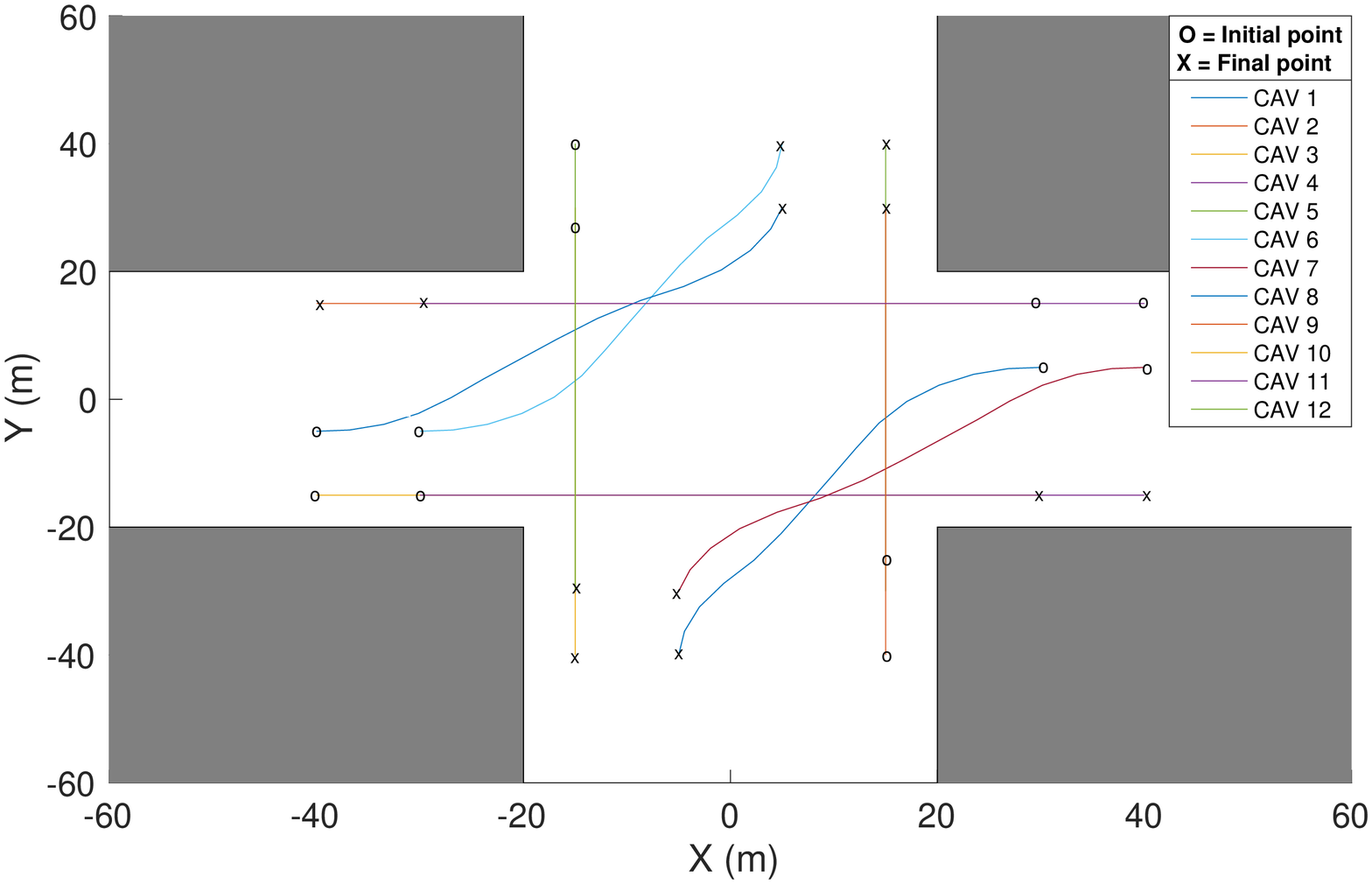}
         \caption{Test scenario one's motion trajectory.}
         \label{Traj_12CAVs}
     \end{subfigure}
     \begin{subfigure}[t]{0.49\textwidth}
         \centering
         \includegraphics[scale=0.33]{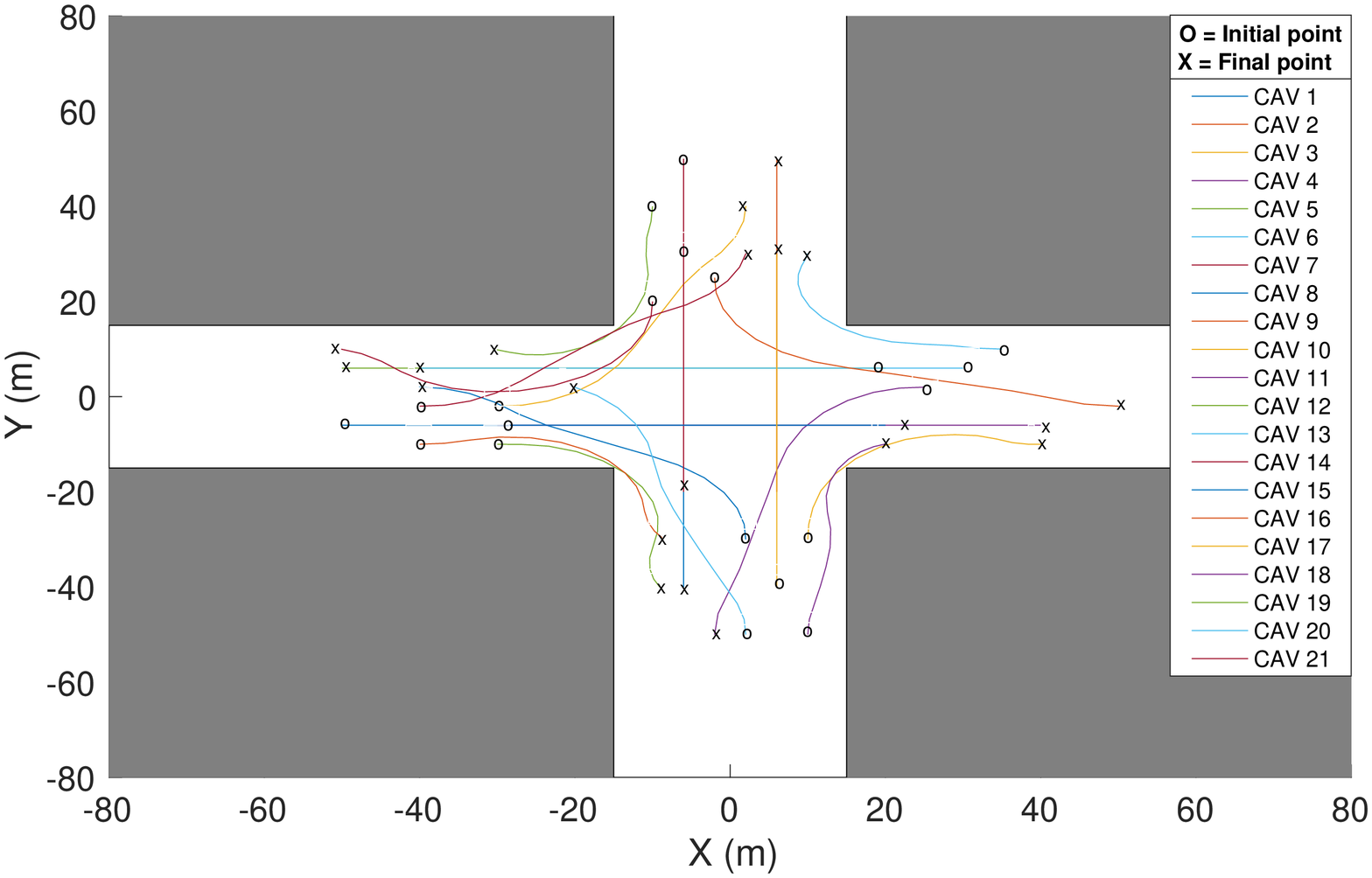}
         \caption{Test scenario two's motion trajectory.}
         \label{21CAVs_Traj}
    \end{subfigure}
        \begin{subfigure}[t]{0.49\textwidth}
         \centering
         \includegraphics[scale=0.33]{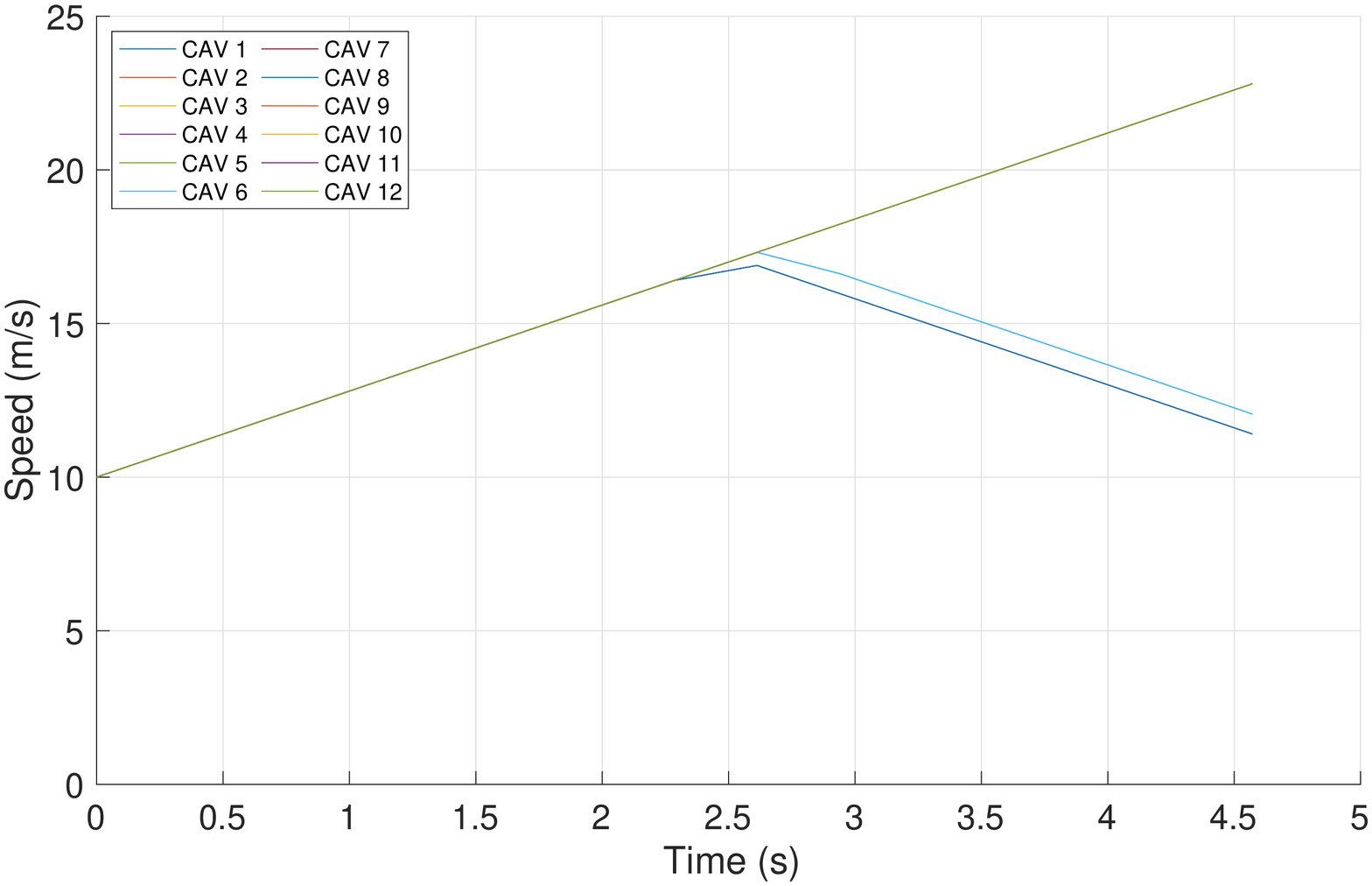}
         \caption{Test scenario one's speed trajectory.}
         \label{scenarioOne_speed}
         \vspace{15pt}
     \end{subfigure}
     \begin{subfigure}[t]{0.49\textwidth}
         \centering
         \includegraphics[scale=0.33]{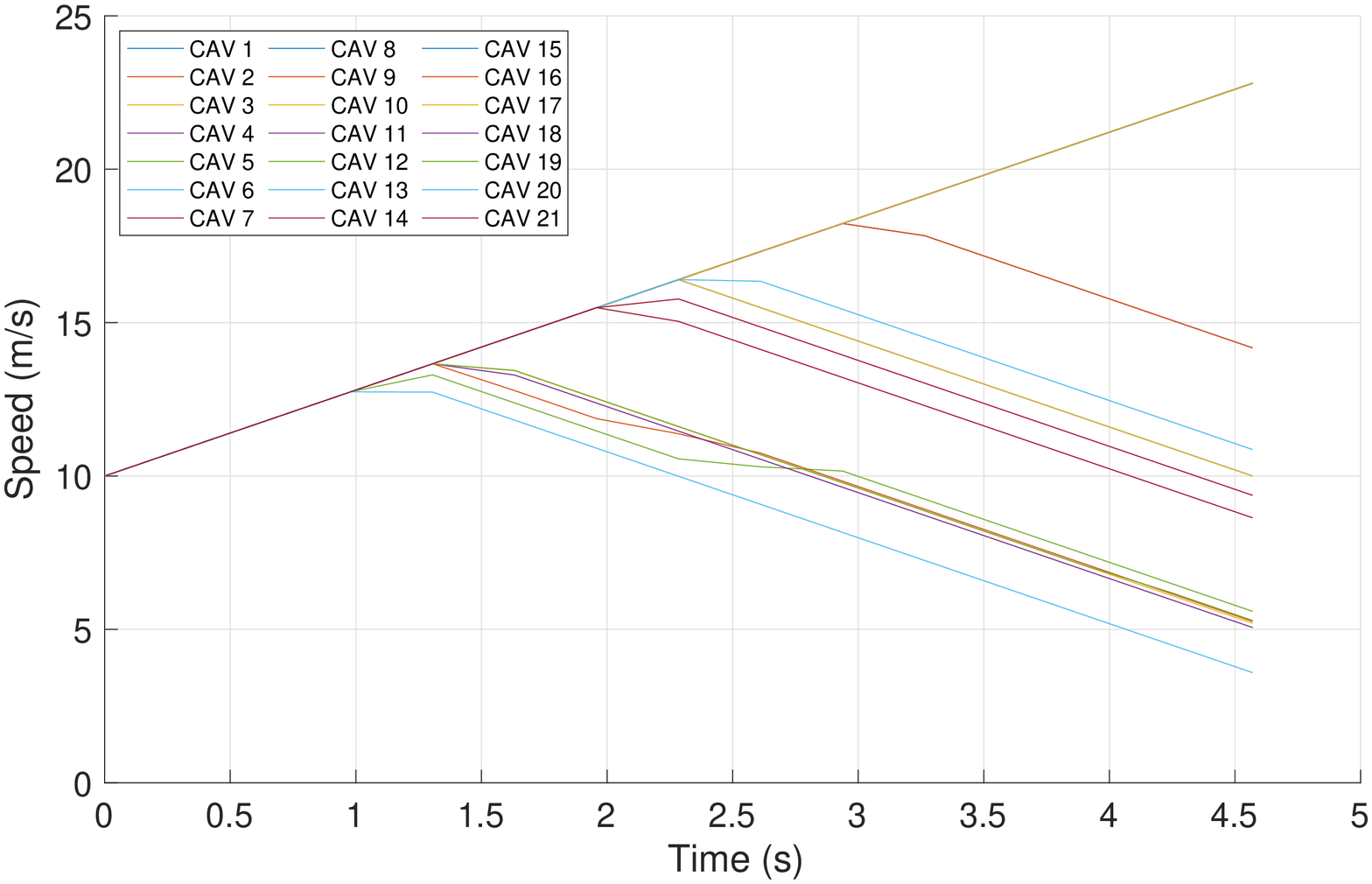}
         \caption{Test scenario two's speed trajectory.}
         \label{scenarioTwo_speed}
    \end{subfigure}
    \vspace{-12pt}
    \caption{The calculated optimal trajectories of motion and speed using the proposed algorithm in test scenario one and test scenario two for 12 and 21 CAVs respectively.}%
    \vspace{-10pt}\label{fig:optimalTrajectories}
\end{figure*}

\subsection{Crossing Time}
\label{crossingTime}
This section compares the minimum crossing time of CAVs that can be achieved by the developed and benchmark algorithms. The acceleration gain $\gamma$ in (\ref{cost}) is set to zero to calculate the minimum-time travelling trajectories of CAVs. In other words, the energy consumption is not considered and CAVs only try to reach destinations as fast as possible, which makes the problem single objective.

Table \ref{benchmark_results} compares crossing time of CAVs when they are controlled by the developed and benchmark algorithms during test scenario one. The table also summarises energy consumption, the travelled distance and average and standard deviation of speed of CAVs. It is worth noting that the crossing time is measured as the time when all the CAVs are already arrived to their destinations. Also, the travelled distance and energy consumption are calculated for all the crossing CAVs. 

As seen in Table \ref{benchmark_results}, crossing time of CAVs when they are controlled by the proposed algorithm is, respectively, up to $65\%$ (for 12 CAVs and in average $52\%$ for all number of crossing CAVs), and $54\%$ less than the case where CAVs are controlled by the reservation-based approach in \cite{malikopoulos2021optimal} and the lane-free method in \cite{li2020autonomous}. This is, of course, in cost of higher energy consumption, as the objective function of the proposed algorithm only considers minimisation of travelling time. The next subsection provides a detail analysis on energy consumption of different approaches, and shows that the proposed algorithm can still achieve significant improvement in crossing time while consuming the same amount of energy as the reservation-based method in \cite{malikopoulos2021optimal}. 

It is also shown in Table \ref{benchmark_results} that, unlike the reservation-based strategy, the resulting crossing time of the proposed algorithm does not change regardless of number of crossing CAVs. There is a similar trend for the average and standard deviation of speed of CAVs when they are controlled by the developed strategy. Also, it is evident from Table \ref{benchmark_results} that the standard deviation of the speed of crossing CAVs when are controlled by the proposed algorithm is mostly less than the case when they are controlled by the reservation based strategy in \cite{malikopoulos2021optimal}. Apparently, the smaller value of standard deviation of speed indicates a less diverge set of speeds (i.e., smoother travel) for the crossing CAVs. 

Table \ref{complex_Ta} shows crossing time, average and standard deviation of speed and travelled distance for different number of CAVs when they are controlled by the proposed strategy in test scenario two. As shown, crossing time of CAVs in test scenario two is the same as the one in test scenario one, and again does not change regardless of the number of CAVs. This determines that the crossing time is not sensitive to the type of scenario and number of CAVs. This is an interesting outcome that shows in lane-free intersections, the crossing time of CAVs is limited by the layout of the junction rather than by the number of passing CAVs, as in traditional signalised intersections.

\begin{figure}[t]
    \centering
    \includegraphics[scale=0.33]{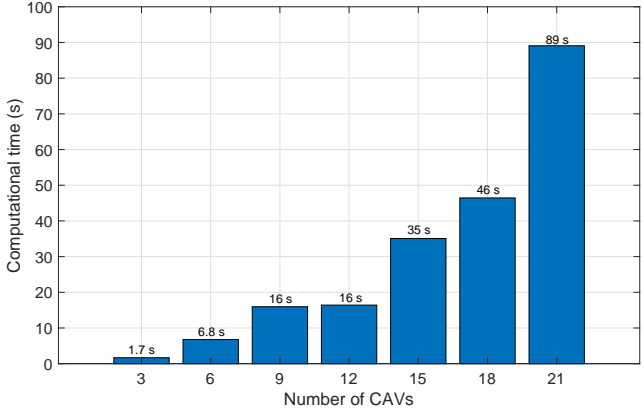}
    \caption{The average computational time of 10 runs of the proposed strategy for different number of CAVs with test scenario two. The standard deviation of the 10 runs for each number of CAVs is less than 0.5\%.}
    \label{ComTime}
\end{figure}

\begin{figure}[t]
    \centering
    \begin{subfigure}[t]{0.49\textwidth}
         \centering
         \includegraphics[scale=0.33]{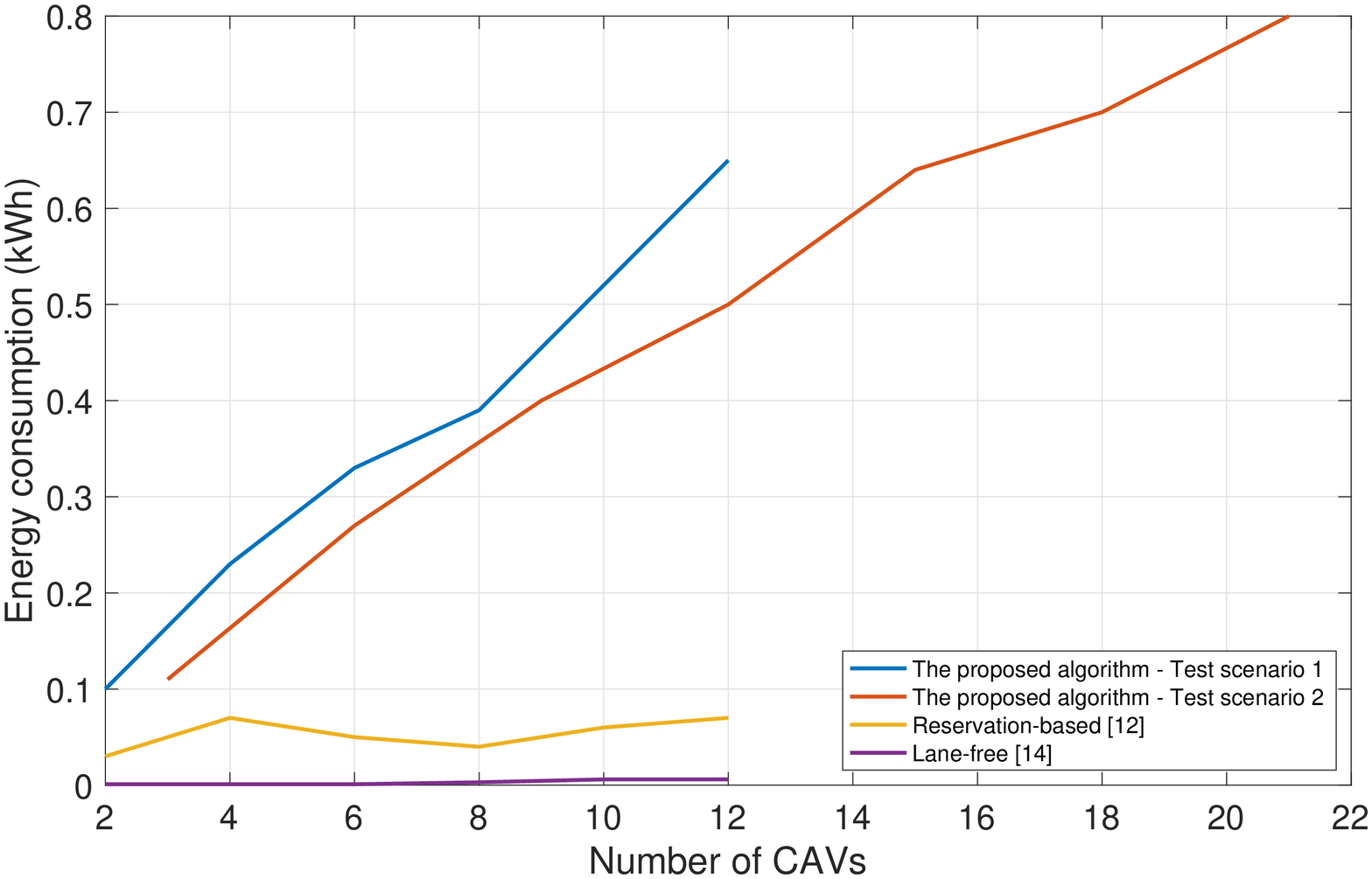}
         \caption{}
         \label{totalEnergy}
    \end{subfigure}
    \begin{subfigure}[t]{0.49\textwidth}
        \centering
        \includegraphics[scale=0.33]{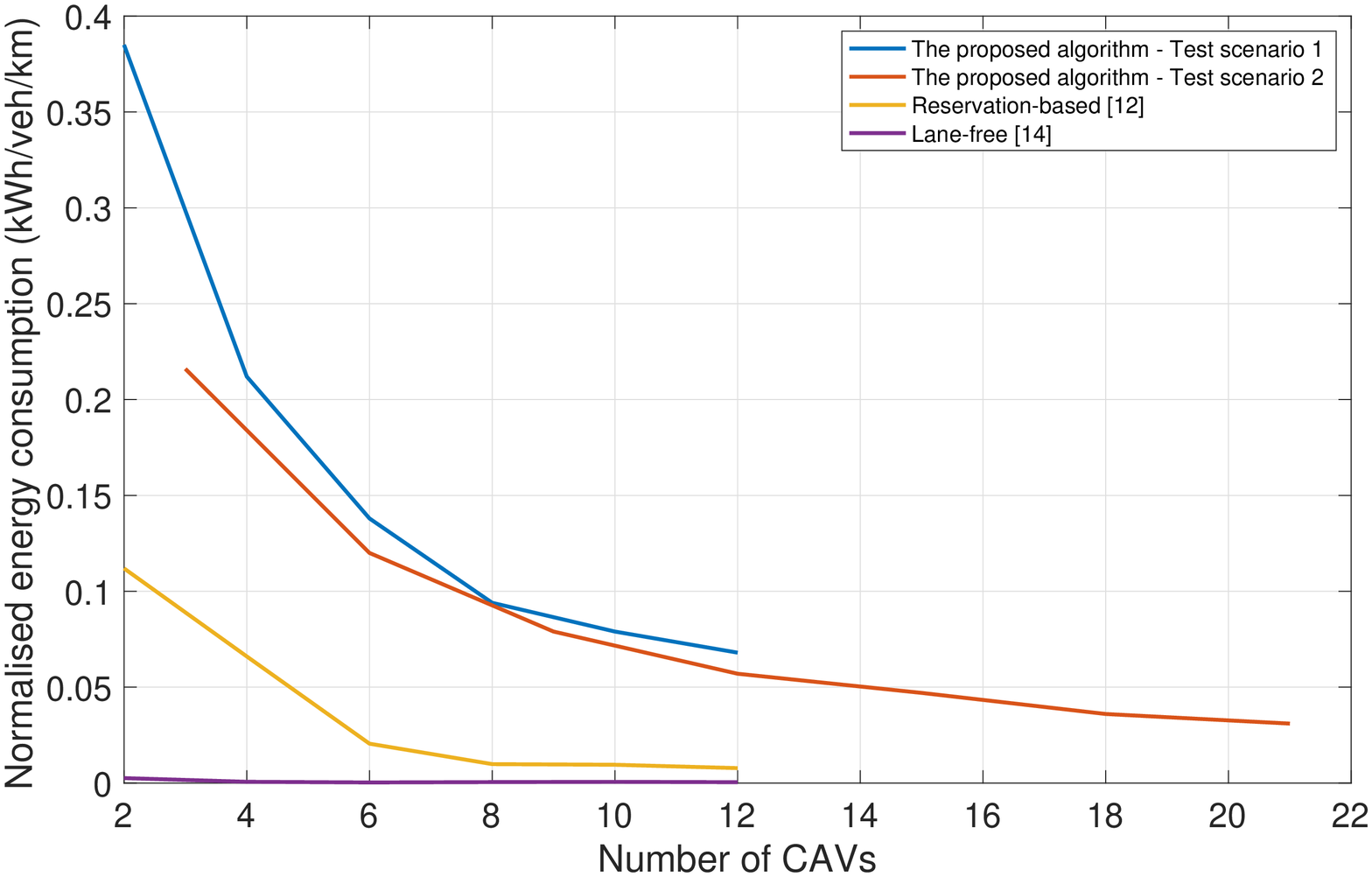}
        \caption{}
        \label{energypervehicleperkm}
        \vspace{15pt}
    \end{subfigure}
    \vspace{-12pt}
    \caption{Comparison of a) total energy consumption b) energy consumed per vehicle per kilometer for different number of vehicles.}%
    \vspace{-10pt}\label{lane_sen}
\end{figure}

In fact, crossing time of CAVs cannot be theoretically smaller than the duration when the CAV with the longest distance from its destination travels its path with the maximum permissible acceleration (i.e., a lower bound of the solution of any lane-free crossing algorithm of intersections). In both test scenarios, the initial speed of CAVs is $10\ (m/s)$, the maximum travelling distance is 70 meters and the maximum acceleration is $3\ (m/s^2)$, and hence the theoretical lower bound of crossing time is 4.27 $s$. The results in Table \ref{benchmark_results} and \ref{complex_Ta} show that the proposed algorithm finds a very close value to this theoretical boundary regardless of type of scenario and number of crossing CAVs. In face, the resulting crossing time can be as close as desired to the theoretical bound in cost of deviation of final point from the desired destination point.

\begin{table}[t]
    \caption{Performance of the proposed lane-free method in test scenario one when the energy consumption is the same as the reservation-based strategy in \cite{malikopoulos2021optimal}.}
    \begin{tabular}{m{3.3cm} m{0.43cm} m{0.43cm} m{0.43cm} m{0.43cm} m{0.43cm} m{0.43cm} m{0.43cm}}
    \toprule
    \textbf{Number of CAVs $\rightarrow$} & \textbf{2} & \textbf{4} & \textbf{6} & \textbf{8} & \textbf{10} & \textbf{12} \\ \hline
    \multicolumn{1}{m{3.3cm}}{Crossing time (s)}  & 5.23 & 5.30 & 5.97   & 6.18   & 6.24   & 6.27\\ 
    \multicolumn{1}{m{3.3cm}}{Average speed (m/s) }  & 13.25 & 13.58 & 11.95   & 11.26   & 11.34   & 11.39\\
    \multicolumn{1}{m{3.3cm}}{Standard deviation of speeds}  & 1.74 & 1.74 & 1.14   & 1.00   & 0.99   & 0.97\\
    \multicolumn{1}{m{3.3cm}}{Energy consumption (kWh) }  & 0.03	&0.07	&0.05	&0.04	&0.06	&0.07\\
    \multicolumn{1}{m{3.3cm}}{Travelled distance (m)}  & 131 & 270 & 401   & 521   & 661   & 801\\
    \bottomrule
    \end{tabular}\label{Energy_considered}
    \vspace{-1pt}
\end{table}

\begin{figure}[t]
    \centering
    \includegraphics[scale=0.33]{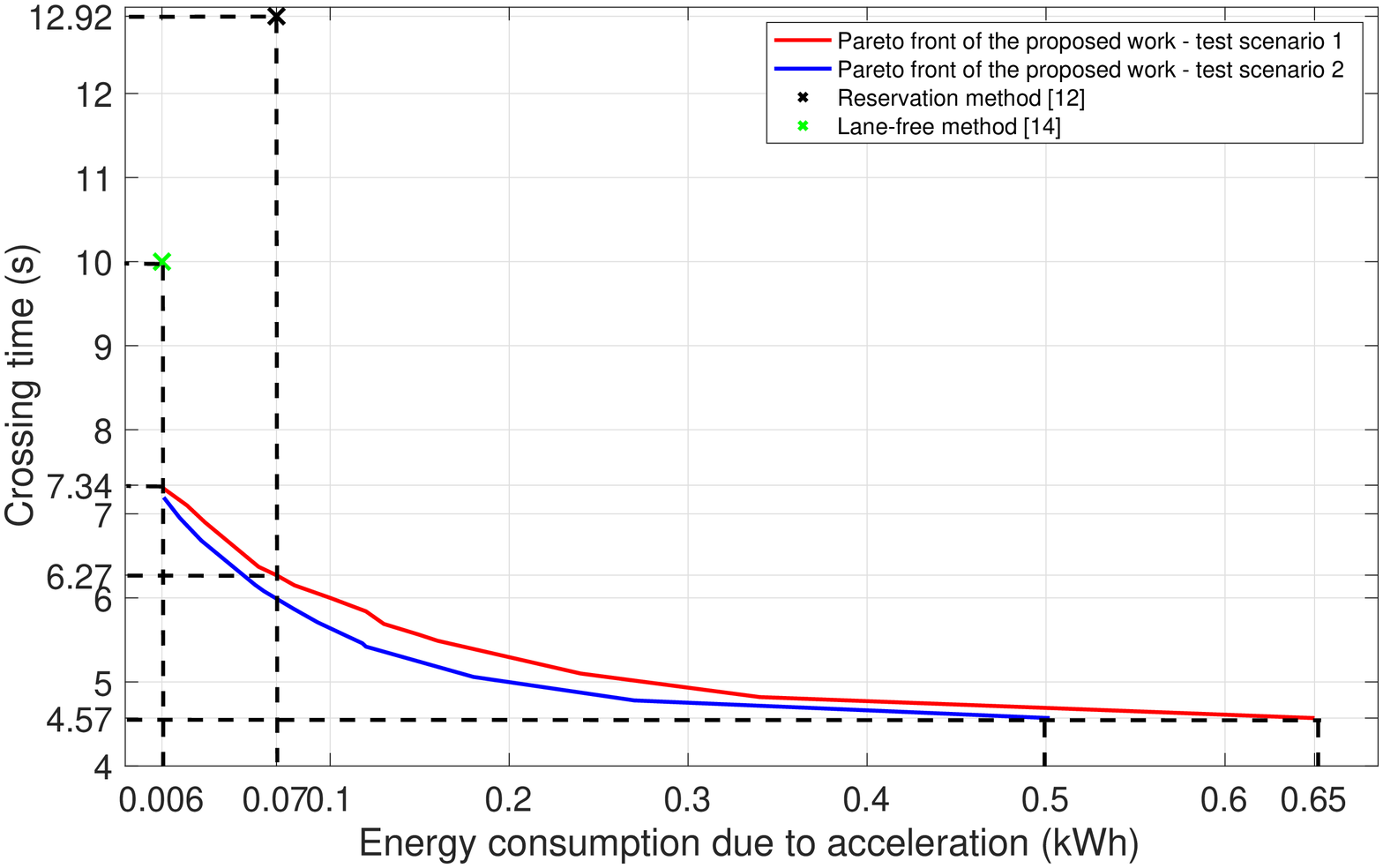}
    \caption{Energy vs crossing time (Pareto front) of 12 CAVs controlled by the proposed strategy as compared to the results by the reservation method in \cite{malikopoulos2021optimal} and the lane-free method in \cite{li2020autonomous}.}
    \label{pareto}
    \vspace{-10pt}
\end{figure}

Fig.~\ref{fig:optimalTrajectories} shows the calculated optimal trajectories and vehicle speeds for both the test scenarios with the maximum number of CAVs (i.e., 12 for test scenario one and 21 for the test scenario two). Fig.s \ref{Traj_12CAVs} and \ref{21CAVs_Traj} illustrate that CAVs move and use opposite lanes freely while avoiding road boundaries. The results are also visualised by a provided video on \color{blue}\url{https://youtu.be/L_aFGkKT38U} \color{black}.

As shown in Fig. \ref{scenarioOne_speed} and \ref{scenarioTwo_speed}, the proposed strategy increases and decreases the speed of CAVs linearly to avoid collisions. The slope of variation (i.e., acceleration and deceleration) is $3~m/s^2$ indicating that it is a bang-bang strategy.

Fig. \ref{ComTime} depicts computational time of the proposed algorithm for different number of CAVs in test scenario two. The computation time of each number of CAV is the average of 10 times of running the scenario. The standard deviations of all the tests are less than $0.5\%$ which is negligible and are not shown. Fig. \ref{ComTime} shows that the computational complexity of the proposed algorithm is of the order of $O(e^{0.13n})$ in terms of number of CAVs $n$. The future work of this study is to reduce the computational time by decentralising the proposed strategy. 

\begin{figure*}[ht]
    \centering
        \begin{subfigure}[t]{0.49\textwidth}
         \centering
         \includegraphics[scale=0.33]{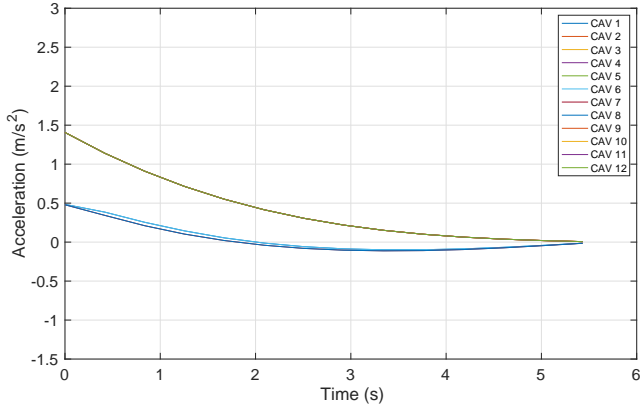}
         \caption{The proposed algorithm's acceleration trajectory.}
         \label{Acceleration_proposed}
         \vspace{15pt}
     \end{subfigure}
     \begin{subfigure}[t]{0.49\textwidth}
         \centering
         \includegraphics[scale=0.33]{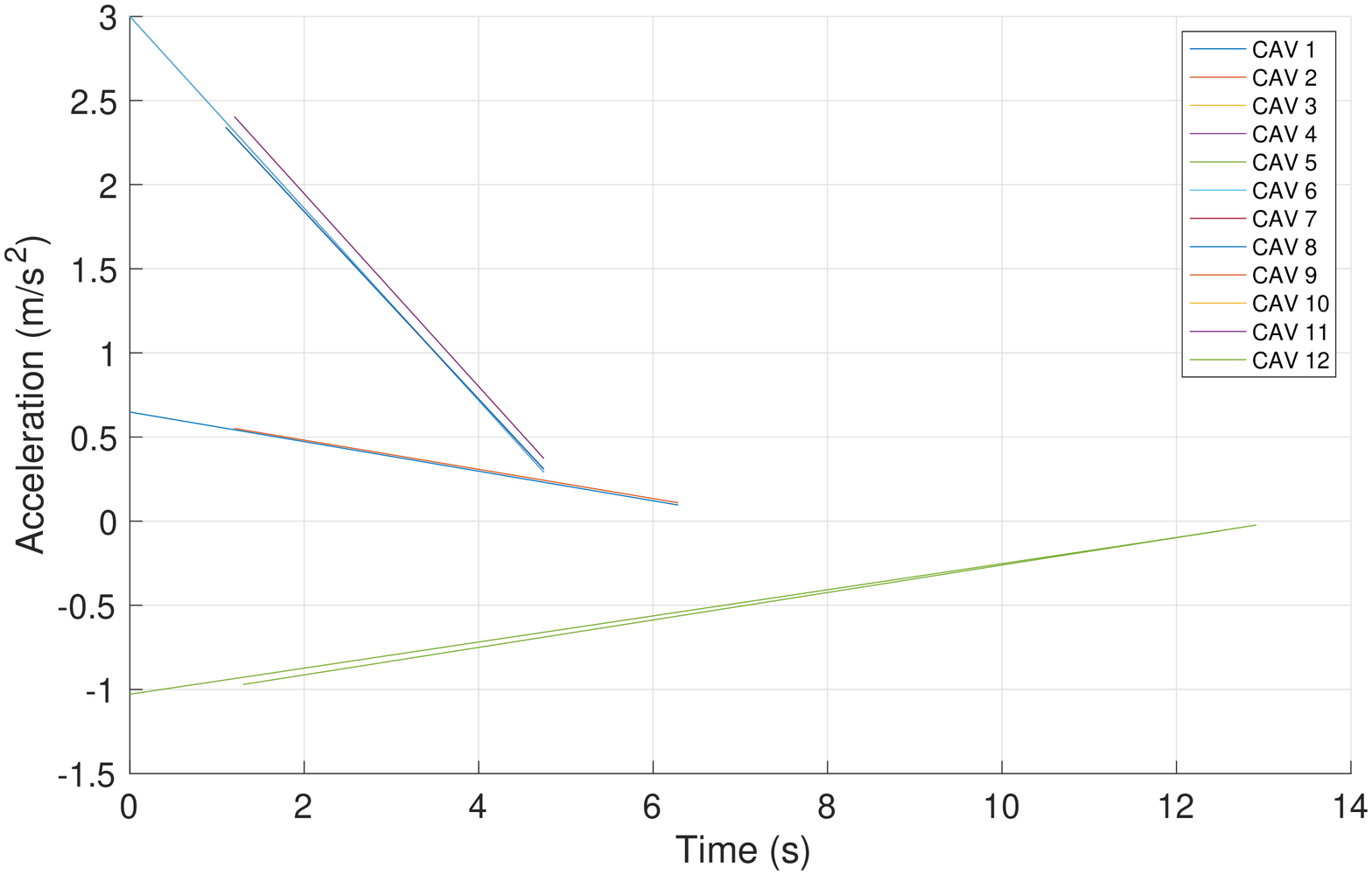}
         \caption{The reservation-based method \cite{malikopoulos2021optimal} acceleration trajectory.}
         \label{Acceleration_reservation}
    \end{subfigure}
    \begin{subfigure}[t]{0.49\textwidth}
         \centering
         \includegraphics[scale=0.33]{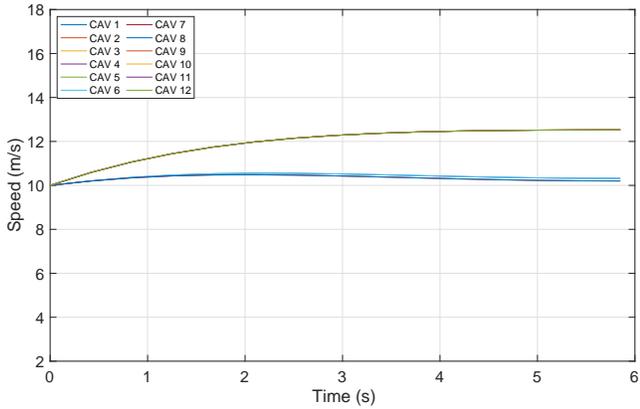}
         \caption{The proposed algorithm's speed trajectory.}
         \label{Speed_proposed}
     \end{subfigure}
     \begin{subfigure}[t]{0.49\textwidth}
         \centering
         \includegraphics[scale=0.33]{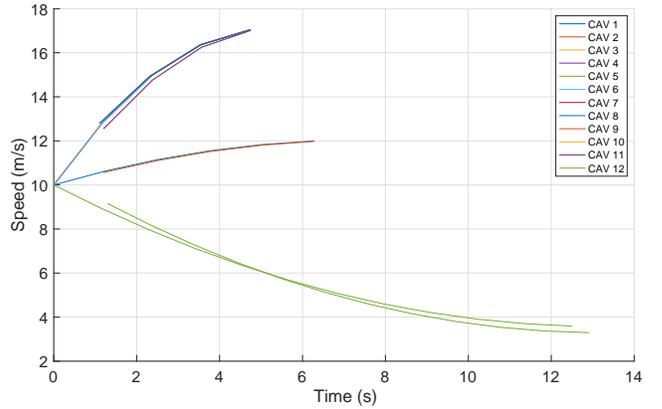}
         \caption{The reservation-based method \cite{malikopoulos2021optimal} speed trajectory.}
         \label{Speed_reservation}
    \end{subfigure}
    \caption{The calculated optimal trajectories of speed and acceleration using the proposed algorithm and reservation-based method \cite{malikopoulos2021optimal} in test scenario one for 12 CAVs, when energy consumption is the same.}%
    \vspace{-10pt}\label{passenger_comfort}
\end{figure*}

\subsection{Energy Consumption}
Fig \ref{totalEnergy} illustrates the total energy consumption of CAVs when the vehicles are controlled by the proposed and benchmark strategies in test scenario one. The figure also shows the total energy consumption of CAVs being controlled by the proposed strategy in test scenario two. The depicted graphs only consider the energy consumption due to acceleration which is calculated as follows:
\begin{equation}
    E_{i} = m\int_{t_{0}}^{t_{f}}a_{i}(t)v_{i}(t)dt \nonumber
\end{equation}
where $E_{i}$ is the energy consumed by each $\text{CAV}_{i}$. 

As seen in Fig. \ref{totalEnergy}, the lane-free method proposed in \cite{li2020autonomous} consumes the least energy, in cost of fixing the crossing time to an unnecessarily large value (i.e., 10~s). The proposed algorithm in this paper, in contrast, consumes more energy than both the benchmark strategies because it is optimised for minimisation of crossing time, as explained in section \ref{crossingTime}.

Moreover, the resulting energy consumption of this proposed strategy linearly increases with respect to the number of crossing CAVs in both the test scenarios. It can also be observed that CAVs consumes more energy in test scenario one than test scenario two because of a longer travelling distance due to diversity of destination of CAVs. Tables \ref{benchmark_results} and \ref{complex_Ta} show the travelled distance of, respectively, test scenario one and two. 

Fig. \ref{energypervehicleperkm}, on the other hand, compares the algorithms in terms of the energy consumption by each vehicle when travels one kilometer. As seen, all the strategies tend to consume less energy per vehicle per kilometer with an increase in the number of CAVs. This is due to the fact that the number of obstacles drops by reducing the number of crossing CAVs, and hence vehicles can accelerate and pass through faster in test scenario one. 

To nullify energy consumption as one of the objectives, and only compare the crossing time of CAVs when they are controlled by the proposed strategy and the reservation-based one in \cite{malikopoulos2021optimal}, the acceleration gain $\gamma$ in (\ref{cost}) is tuned such that energy consumption of CAVs in both cases becomes the same. Table \ref{Energy_considered} shows the resulting performance of the proposed algorithm. As compared to the results of the reservation-based method in Table \ref{benchmark_results}, the proposed algorithm reduces the crossing time up to 52\% (average of 40\%) when consumes the same amount of energy. Moreover, whilst the average speed of CAVs is almost similar for both the strategies, the standard deviation of speed of CAVs controlled by the proposed algorithm is much lower. This indicates that CAVs controlled by the proposed algorithm travel with similar speed, whilst some of the CAVs being controlled by the reservation-based method travel with a much higher or lower speed than the others. 

Fig. \ref{pareto} depicts that the proposed algorithm finds the Pareto front of all the crossing solutions of 12 CAVs for different values of acceleration gain $\gamma$. As seen, the proposed strategy can achieve shorter crossing time than both the reservation-based method \cite{malikopoulos2021optimal} and lane-free method \cite{li2020autonomous} while consuming the same amount of energy. Moreover, the resulting crossing time of CAVs can be as close as possible to its theoretical lower bound.

\subsection{Passenger Comfort}
Fig. \ref{passenger_comfort} compares the calculate optimal speed and acceleration (or deceleration) trajectories by the proposed algorithm and reservation-based method in \cite{malikopoulos2021optimal} for 12 CAVs in test scenario one, when energy consumption is the same. 

As shown in Fig.~\ref{Acceleration_proposed}, the maximum deceleration of CAVs when they are controlled by the proposed strategy is $1.4~m/s^2$ which is much less than the maximum permissible value of 3 $(m/s^2)$. Moreover, the acceleration of all CAVs converges to zero at their destinations. Fig. \ref{Acceleration_reservation}, on the other hand, shows that some of the CAVs controlled by the reservation-based algorithm in \cite{malikopoulos2021optimal} decelerate with the maximum permissible value, which is not converged to zero. 

The maximum jerk of both algorithms is around 0.6 $m/s^3$, however, whilst jerks of CAVs controlled by the proposed strategy converges to zero, the passengers feel an uncomfortably constant jerk during the crossing time when CAVs are controlled by the reservation-based method in \cite{malikopoulos2021optimal}.

Fig.~\ref{Speed_proposed} as compared to Fig.~\ref{Speed_reservation} shows that all the vehicles travel within a much narrower range of speeds when they are controlled by the proposed algorithm. This means their passengers will experience similar feeling of speed, unlike the passengers of CAVs controlled by the reservation-based method in \cite{malikopoulos2021optimal} who experience a diverged range of speeds (along with larger constantly applied jerk and higher acceleration/deceleration).

\section{Conclusion}\label{conclusion}

This study formulates the lane-free crossing of CAVs through intersections as an optimal control problem that minimises the overall crossing time and energy consumption of CAVs while avoiding obstacles. The proposed formulation employs dual problem theory to substitute the non-differentaiable and non-convex constraints of collision avoidance with the dual problem of a corresponding sufficient condition. 

The resulting smoothed OCP is then solved by CasADi to generate a strategy for safely cross of multiple CAVs through a junction within the minimum time. It is shown that the proposed strategy is capable of significantly reducing the crossing time as compared to the state-of-the-art reservation-based or other lane-free strategies, whilst consuming similar energy.   

The presented results also show that the crossing time of CAVs controlled by the proposed algorithm is very close to its theoretical limit, and only relies on the layout of intersection and is independent of the number of crossing CAVs or their manoeuvres. This makes the results of the proposed algorithm a suitable benchmark to evaluate the performance of other control strategies of the CAVs crossing intersections.

Computational complexity of the proposed algorithm is currently of the order of $O(e^{0.13n})$, where $n$ is the number of CAVs passing through the intersection, that has to be reduced for real-time implementation. This will be added as a future work.



\bibliographystyle{IEEEtran}
{\footnotesize
\bibliography{References}}

\vspace{-25pt}

\begin{IEEEbiographynophoto}{Mahdi Amouzadi}
received BEng degrees in electrical and electronic engineering from, the University of Sussex, Brighton, UK, in 2019. He is currently pursing a Ph.D. degree in the Smart Vehicles Control Laboratory (SVeCLab) in the department of Engineering and Informatics at the University of Sussex. His research interests lies at the intersection of control and optimization of connected autonomous vehicles (CAVs). He is focused on designing safe and efficient path planning algorithms for congested areas such as intersections. 
\end{IEEEbiographynophoto}

\vspace{-25pt}

\begin{IEEEbiographynophoto}{Mobolaji Olawumi Orisatoki} received his B.Sc. degree in computer science from the University of Greenwich, in 2006, M.Sc. degree from Royal Holloway, University of London, in 2012, and PGCE from Institute of Education-University College London, in 2013. He is currently pursuing a Ph.D. degree in the Smart Vehicles Control Laboratory (SVeCLab) in the Department of Engineering and Design, University of Sussex, UK. He has taught in different colleges across South and East London. His research interests include system optimization and control, system dynamics, multi-agent systems, information entropy and mapping. His focused on designing an algorithm to map of a non-convex area, maximizing information with travelling distance.
\end{IEEEbiographynophoto}

\vspace{-25pt}

\begin{IEEEbiographynophoto}{Arash M. Dizqah (M'12)}
received MEng and MSc degrees in electrical engineering from, respectively, Sharif and KN Toosi Universities of Technology, Tehran, Iran, in 1998 and 2001, and his PhD degree in control engineering from Northumbria University, UK, in 2014.
He was a research fellow with the University of Surrey, UK and he is currently a Senior Lecturer in mechanical engineering at the University of Sussex. His research interests lie in control and optimisation with applications in vehicles and robotics. He is particularly interested in the real-time implementation of distributed nonlinear optimisation-based controllers for connected and autonomous vehicles (CAVs) and for informative path planning with a team of robots.
Arash is the director of the Smart Vehicles Control Laboratory (SVeCLab), focusing on advanced control strategies for connected and autonomous vehicles, swarm robotics and energy management strategies.
\end{IEEEbiographynophoto}

\vfill

\end{document}